# The Structure-Behavior Coalescence Method

## --Toward a Unified View of the Software System in Model-Driven Engineering

William S. Chao

*Abstract*—In Model-Driven Engineering (MDE), the Unified Modeling Language (UML) 2.0 specification includes a metamodel that defines the language concepts and a user model that defines how the language concepts are represented. In UML 2.0, an important use of metamodel is to provide an integrated semantic framework that every diagram in the user model can be projected as a view of the metamodel. However, most existing UML 2.0 metamodels lack the ability to serve as the basis for unifying different views of a software system. To overcome the shortcomings of the current UML 2.0 metamodel approaches, we developed Operation-Based Multi-Queue Structure-Behavior Coalescence Process Algebra (O-M-SBC-PA), which provides an integrated semantic framework that is able to integrate structural and behavioral constructs. Using O-M-SBC-PA as the metamodel of UML 2.0, each diagram in the user model can be projected as a view of the MDE metamodel.

*Index Terms*—Model-Driven Engineering, Unified Modeling Language, Metamodel, User Model, Operation-Based Multi-Queue Structure-Behavior Coalescence Process Algebra, Integrated Semantic Framework

## I. INTRODUCTION

As a software modeling language for model-driven engineering (MDE) applications [1], the Unified Modeling Language (UML) 2.0 [2]-[3] defines a set of language concepts that are used to model the structure and behavior of a software system. The UML 2.0 concepts include (a) an abstract syntax that defines the language concepts and is described by a metamodel, and (b) a concrete syntax, or notation, that defines how the language concepts are represented and is described by a user model [4]-[5].

Since UML 2.0 is a multi-diagram approach, there are always some inconsistencies between different diagrams in the user model [6]-[8]. To ensure and check the consistency, a metamodel that defines the abstract syntax of a modeling language needs to provide an integrated semantic framework for defining consistency rules to impose constraints on the software structure or behavior constructs. It is hoped that through this integrated semantic framework, each diagram in the user model can be projected as a view of the metamodel.

Unfortunately, most current UML 2.0 metamodels are not able to project each diagram in the user model as a view of the metamodel. In this paper, we developed Operation-Based Multi-Queue Structure-Behavior Coalescence Process Algebra (O-M-SBC-PA) [9] as a metamodel of UML 2.0. In O-M-SBC-PA, each diagram in the user model will be projected as a view of the metamodel. Therefore, we claim that O-M-SBC-PA genuinely provides a coalesced semantic framework to ensure model consistency for UML 2.0.

The remainder of this paper is arranged as follows. Section 2 deals with the current UML 2.0 metamodel study. O-M-SBC-PA as a metamodel for UML 2.0 is detailed in Section 3. After detailing O-M-SBC-PA, we will validate the SBC method with a case study in Section 4. Conclusions of this paper are in Section 5.

## II. RELATED STUDIES

. A metamodel of UML 2.0 is used to describe the concepts in the language, their characteristics, and interrelationships. This is sometimes called the abstract syntax of the language, and is distinct from the concrete syntax that specifies the user model for the language. A significant usage of the metamodel is to ensure model consistency between different diagrams in the user model.

The Object Management Group (OMG) defines a language for representing metamodes, called Meta Object Facility (MOF) that is used to define UML, SysML and other metamodels. Several mechanisms are used in MOF, such as Object Constraint Language (OCL) [10], Foundational UML (fUML) [11], The Action Language for Foundational UML (Alf) [12], Process Specification Language (PSL) [13], to name a few.

The Object Constraint Language is a precise text language that provides constraint on the structure (i.e., objects) to ensure consistency of the user model [10]. However, not each diagram in the user model can be projected as a view of the OCL metamodel because the OCL fails to provide an integrated semantic framework. Therefore, the OCL metamodel can only ensure part of (not all) user model consistency.

The Foundational UML is a subset of the standard UML for which a standard execution constraint language, PSL, is used to define the semantics of the execution model [11]. Although fUML provides constraint on the behavior (i.e., activities) to make the model executable, it fails to unify the software structural constructs with the software behavioral constructs. Not being able to provide an integrated semantic framework, the Foundational UML can not project each diagram in the user model as a view of the fUML metamodel.

The Action Language for Foundational UML (Alf) is a complementary specification to Foundational UML [12]. The key use of Alf is to act as the notation for specifying executable software behaviors in UML, for example, methods for object operations, the behavior of an object, or transition effects on state machines. Like fUML, Alf also fails to provide an integrated semantic framework to unify the software structural constructs with the software behavioral constructs. Therefore, the Alf is not able to project each diagram in the user model as a view of the Alf metamodel.

In order to overcome the shortcomings of the current UML 2.0 metamodel approaches, we need to develop an integrated semantic framework that is able to unify structural constructs with behavioral constructs. Operation-Based Multi-Queue Structure-Behavior Coalescence Process Algebra (O-M-SBC-PA) is such a candidate. In O-M-SBC-PA, the software structural and software behavioral constructs are unified. Using O-M-SBC-PA as a metamodel for UML 2.0, each diagram in the user model can be projected as a view of the MDE metamodel.

### III. METHOD OF OPERATION-BASED MULTI-QUEUE SBC PROCESS ALGEBRA

#### A. Operation-Based Value-Passing Interactions

The object is the fundamental modular unit for describing software structure in UML [1]-[3]. An operation represents a procedure, method, or function that an object performs when a caller calls it. Each operation defines a set of parameters that describes the arguments passed in with the request, or passed back out once a request has been handled. An operation (can be extended to operation call or operation return) signature is a combination of its name along with parameters as follows:

<operation name> ( )

The parameters in the parameter list represent the inputs or outputs of the operation. Each parameter in the list is displayed with the following format:

<direction> : Parameter direction may be in, out, or inout. We formally describe the "operation call or operation return signature" as a relation $L \subseteq \Lambda \times \Theta$ where $\Lambda$ is a set of "operation names" and $\Theta$ is a set of "parameter lists".

An interaction represents an indivisible and instantaneous handshake or rendezvous between the caller agent (either external environment's actor or object) and the callee agent (object) [9]. In the operation-based value-passing interaction approach as shown in Figure 1, the caller agent interacts with the callee object through the operation call or operation return interaction. In the figure, "getPastDueBalance(in studentId)" is an operation call signature and "getPastDueBalance(out PastDueBalance)" is an operation return signature. The operation call signature and its corresponding operation return signature can be merged into an operation signature.

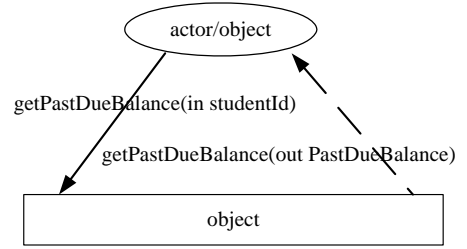

Figure 1. Operation-based Value-passing Interaction

We formally describe the "operation-based value-passing interaction" as a relation $\Delta \subseteq N \times \Xi \times \Lambda \times \Theta \times \Gamma$, where $N$ is a set of "operation call or operation return tags" and $\Xi$ is a set of "external environment's actors or objects" and $\Lambda$ is a set of "operation names" and $\Theta$ is a set of "parameter lists" and $\Gamma$ is a set of "objects".

#### B. SBC Interaction Transition Graphs in O-M-SBC-PA

In O-M-SBC-PA, we use the SBC interaction transition graph (ITG) as a single diagram to specify the semantics of a software system. The SBC interaction transition graph is a labelled transition system (LTS) [14]. Overall, the SBC interaction transition graph provides an integrated semantic framework to unify structural constructs with behavioral constructs. In O-M-SBC-PA, each state is regarded as a process. The notion of a SBC interaction transition graph is defined as follows.

**DEFINITION** (INTERACTION TRANSITION GRAPH)
A SBC interaction transition graph $ITG = (\Psi, s_0, N, \Xi, \Lambda, \Theta, \Gamma, ITGR)$ consists of

- *a finite set $\Psi$ of states*,
- *an initial state $s_0 \in \Psi$*,
- *a finite set $N$ of operation call or operation return tags*,
- *a finite set $\Xi$ of external environment's actors or objects*,
- *a finite set $\Lambda$ of operation names*,
- *a finite set $\Theta$ of parameter lists*,
- *a finite set $\Gamma$ of objects*,

- *a transition relation* $ITGR \subseteq \Psi_1 \times N \times \Xi \times \Lambda \times \Theta \times \Gamma \times \Psi_2$, where $(s_j, n, \rho, op, p, b, s_k) \in ITGR$ is denoted by $s_j \xrightarrow{n, \rho, op, p, b} s_k$.

We can draw a diagram to represent the SBC interaction transition graph. Figure 2 shows the diagram of the SBC interaction transition graph $ITG_{01}$. In the diagrammed SBC interaction transition graph, the state is represented by a circle containing its name; the transition from the source state to the target state is represented by an arrow and labelled with an interaction; the initial state (for example, "$s_1$") is the target state of the transition that has no source state. In a state, if multiple transitions to be triggered are met, the choice of trigger will be arbitrary and fair.

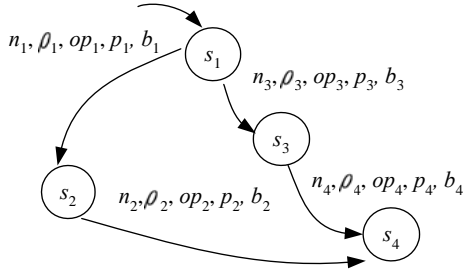

Figure 2. Diagram of the Transition Graph $ITG_{01}$

We can also list the relationships that represent the SBC interaction transition graph. Table 1 shows the transition relation $ITGR_{01}$ of the SBC interaction transition graph $ITG_{01}$.

TABLE 1. Relation $ITGR_{01}$ of the SBC interaction transition graph $ITG_{01}$

| $\Psi_1$ | $N$ | $\Xi$ | $\Lambda$ | $\Theta$ | $\Gamma$ | $\Psi_2$ |
|---|---|---|---|---|---|---|
| $s_1$ | $n_1$ | $\rho_1$ | $op_1$ | $p_1$ | $b_1$ | $s_2$ |
| $s_2$ | $n_2$ | $\rho_2$ | $op_2$ | $p_2$ | $b_2$ | $s_4$ |
| $s_1$ | $n_3$ | $\rho_3$ | $op_3$ | $p_3$ | $b_3$ | $s_3$ |
| $s_3$ | $n_4$ | $\rho_4$ | $op_4$ | $p_4$ | $b_4$ | $s_4$ |

In order to reduce the complexity of the SBC interaction transition graph, we shall introduce an orthogonal composite state. An orthogonal composite state in the SBC interaction transition graph may have many regions, which may each contain substates. These regions are orthogonal to each other. When an orthogonal composite state is active, each region has its own active state that is independent of the others and any incoming interaction is independently analyzed within each region. We use $ITG_1 \| ITG_2 \| ITG_3 \| \ldots \| ITG_m$ to represent an orthogonal composite state, which means the composition of $ITG_1$, $ITG_2$, $ITG_3$,…, and $ITG_m$.

C. *SBC Interaction Transition Graph of a Software System*

In O-M-SBC-PA, the SBC interaction transition graph of a software system $ITG_{system}$ is defined as $\|_{i=1, m} ITG_i$ or $ITG_1 \| ITG_2 \| ITG_3 \ldots \| ITG_m$. Each SBC interaction transition graph $TG_i$ is represented by a transition relation $ITGR_i \subseteq \Psi_1 \times N \times \Xi \times \Lambda \times \Theta \times \Gamma \times \Psi_2$, where $(s_{ij}, n, \rho, op, p, b, s_{ik}) \in ITGR_i$ is written as $s_{ij} \xrightarrow{n, \rho, op, p, b} s_{ik}$. The SBC interaction transition graph of a software system $ITG_{system}$ is represented by the transition relation $ITGR_{system}$ which is defined as $\|_{i=1, m} ITGR_i$ or $ITGR_1 \| ITGR_2 \| ITGR_3 \ldots \| ITGR_m$.

We can draw a diagram to represent the SBC interaction transition graph of a software system. Figure 3 shows the diagram of the SBC interaction transition graph $ITG_{system}$.

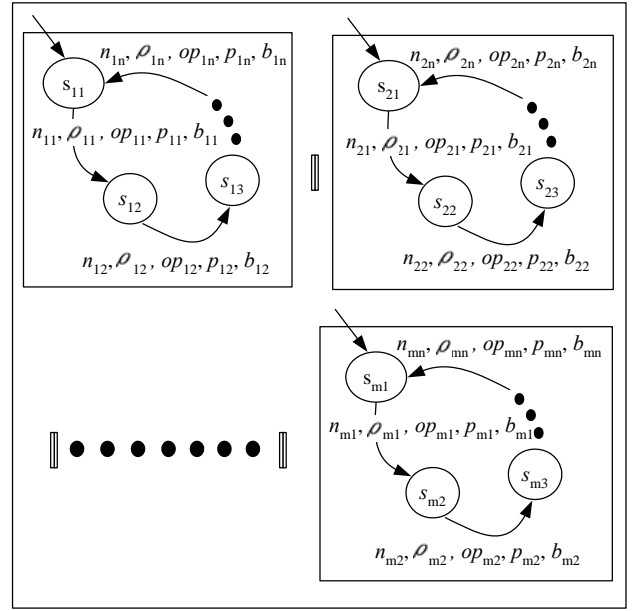

Figure 3. Transition Graph $ITG_{system}$

We can also list the relationships that represent the SBC interaction transition graph of a software system. Table 2 shows the transition relation $ITGR_{system}$ of the SBC interaction transition graph $ITG_{system}$.

TABLE 2. Relation $ITGR_{system}$

| $\Psi_1$ | $N$ | $\Xi$ | $\Lambda$ | $\Theta$ | $\Gamma$ | $\Psi_2$ |
|---|---|---|---|---|---|---|
| $s_{11}$ | $n_{11}$ | $\rho_{11}$ | $op_{11}$ | $p_{11}$ | $b_{11}$ | $s_{12}$ |
| $s_{12}$ | $n_{12}$ | $\rho_{12}$ | $op_{12}$ | $p_{12}$ | $b_{12}$ | $s_{13}$ |
| • | • | • | • | • | • | • |
| $s_{1n}$ | $n_{1n}$ | $\rho_{1n}$ | $op_{1n}$ | $p_{1n}$ | $b_{1n}$ | $s_{11}$ |

| $\Psi_1$ | $N$ | $\Xi$ | $\Lambda$ | $\Theta$ | $\Gamma$ | $\Psi_2$ |
|---|---|---|---|---|---|---|
| $s_{21}$ | $n_{21}$ | $\rho_{21}$ | $op_{21}$ | $p_{21}$ | $b_{21}$ | $s_{22}$ |
| $s_{22}$ | $n_{22}$ | $\rho_{22}$ | $op_{22}$ | $p_{22}$ | $b_{22}$ | $s_{23}$ |
| • | • | • | • | • | • | • |
| $s_{2n}$ | $n_{2n}$ | $\rho_{2n}$ | $op_{2n}$ | $p_{2n}$ | $b_{2n}$ | $s_{21}$ |

‖ • • • • ‖

| $\Psi_1$ | $N$ | $\Xi$ | $\Lambda$ | $\Theta$ | $\Gamma$ | $\Psi_2$ |
|---|---|---|---|---|---|---|
| $s_{m1}$ | $n_{m1}$ | $\rho_{m1}$ | $op_{m1}$ | $p_{m1}$ | $b_{m1}$ | $s_{m2}$ |
| $s_{m2}$ | $n_{m2}$ | $\rho_{m2}$ | $op_{m2}$ | $p_{m2}$ | $b_{m2}$ | $s_{m3}$ |
| • | • | • | • | • | • | • |
| $s_{mn}$ | $n_{mn}$ | $\rho_{mn}$ | $op_{mn}$ | $p_{mn}$ | $b_{mn}$ | $s_{m1}$ |

*D. Projecting the UML Class Diagram from the SBC Interaction Transition Graph*

In UML 2.0, the class diagram is a static structure diagram that describes the structure of a software system by showing the classes of the software system, their attributes, operations (or methods), and relationships between objects. The notion of a UML 2.0 class diagram is defined as follows.

**DEFINITION** (CLASS DIAGRAM) A UML class diagram $ClsD = (C, \Lambda, \Theta, ClsDR)$ consists of

- *a finite set $C$ of classes*,
- *a finite set $\Lambda$ of operation names*,
- *a finite set $\Theta$ of parameter lists*,
- *a relation $ClsDR \subseteq C \times \Lambda \times \Theta$, where $(c, op, p) \in ClsDR$*.

The UML 2.0 class diagram of a software system $ClsD_{system}$ is represented by a relation $ClsDR_{system} \subseteq C \times \Lambda \times \Theta$, where $(c, op, p) \in ClsDR_{system}$, as shown in Table 3.

TABLE 3. Relatiopn $ClsDR_{system}$

| $C$ | $\Lambda$ | $\Theta$ |
|---|---|---|
| $c_1$ | $op_1$ | $p_1$ |
| $c_1$ | $op_2$ | $p_2$ |
| $c_1$ | $op_3$ | $p_3$ |
| $c_4$ | $op_4$ | $p_4$ |
| • | • | • |
| $c_m$ | $op_m$ | $p_m$ |

The algorithm used to project the ClsD relation $ClsDR_{system} \subseteq C \times \Lambda \times \Theta$ from the ITG relation $ITGR_{system} \subseteq S_1 \times N \times \Xi \times \Lambda \times \Theta \times \Gamma \times S_2$ is as follows.

**ALGORITHM 1** (Projecting $ClsDR_{system}$ from $ITGR_{system}$)

**For** i = 1, m **Loop**

SELECT DISTINCT $\Gamma$, $\Lambda$, $\Theta$ INTO $ClsDR_i$ (C, $\Lambda$, $\Theta$) FROM $ITGR_i$ WHERE $N$ = 'OPERATION_CALL';

SELECT DISTINCT $\Gamma$, $\Lambda$, $\Theta$ INTO $ClsDR\_RETURN_i$ (C, $\Lambda$, $\Theta$) FROM $ITGR_i$ WHERE $N$ = 'OPERATION_RETURN';

UPDATE $ClsDR_i$ SET $\Lambda$, $\Theta$ = MERGE (Operation Call Signature, Operation Return Signature) WHERE there exists corresponding Operation Return Signature in $ClsDR\_RETURN_i$;

INSERT INTO $ClsDR_{i\sim m}$ (C, $\Lambda$, $\Theta$) SELECT * FROM $ClsDR_i$;

**End Loop;**

SELECT DISTINCT * INTO $ClsDR_{system}$ FROM $ClsDR_{i\sim m}$

**END ALGORITHM**

Once we have the ClsD relation $ClsDR_{system}$, it is easy to get a UML 2.0 class diagram of the software system.

*E. Projecting the UML State Diagram from the SBC Interaction Transition Graph*

In UML 2.0, the state diagram represents behavior of a software system in terms of its transition between states triggered by actions. The notion of a UML 2.0 state diagram is defined as follows.

**DEFINITION** (STATE DIAGRAM) A UML state diagram $StD = (\Psi, N, \Lambda, StDR)$ consists of

- *a finite set $\Psi$ of states*,
- *a finite set $N$ of operation call or operation return tags*,
- *a finite set $\Lambda$ of operation names*,
- *a relation $StDR \subseteq \Psi_1 \times N \times \Lambda \times \Psi_2$, where $(s_j, n, op, s_k) \in StDR$ is denoted by $s_j \xrightarrow{n,\ op} s_k$*.

The UML 2.0 state diagram of a software system $StD_{system}$ is defined as $\|_{i=1,m} StD_i$ or $StD_1 \| StD_2 \| \ldots \| StD_m$. Each state diagram $StD_i$ is represented by a relation $StDR_i \subseteq \Psi_1 \times N \times \Lambda \times \Psi_2$, where $(s_{ij}, n, op, s_{ik}) \in StDR_i$ is denoted by $s_{ij} \xrightarrow{n, op} s_{ik}$. The state diagram of a software system $StD_{system}$ is represented by the relation $StDR_{system}$ which is defined as $\|_{i=1,m} StDR_i$ or $StDR_1 \| StDR_2 \| \ldots \| StDR_m$, as shown in Table 4.

TABLE 4. Relation $StDR_{system}$

| $\Psi_1$ | $N$ | $\Lambda$ | $\Psi_2$ |
|---|---|---|---|
| $s_{11}$ | $n_{11}$ | $op_{11}$ | $s_{12}$ |
| $s_{12}$ | $n_{12}$ | $op_{12}$ | $s_{13}$ |
| • | • | • | • |
| $s_{1n}$ | $n_{1n}$ | $op_{1n}$ | $s_{11}$ |

$\|$

| $\Psi_1$ | $N$ | $\Lambda$ | $\Psi_2$ |
|---|---|---|---|
| $s_{21}$ | $n_{21}$ | $op_{21}$ | $s_{22}$ |
| $s_{22}$ | $n_{22}$ | $op_{22}$ | $s_{23}$ |
| • | • | • | • |
| $s_{2n}$ | $n_{2n}$ | $op_{2n}$ | $s_{21}$ |

$\| \bullet \bullet \bullet \bullet \bullet \|$

| $\Psi_1$ | $N$ | $\Lambda$ | $\Psi_2$ |
|---|---|---|---|
| $s_{m1}$ | $n_{m1}$ | $op_{m1}$ | $s_{m2}$ |
| $s_{m2}$ | $n_{m2}$ | $op_{m2}$ | $s_{m3}$ |
| • | • | • | • |
| $s_{mn}$ | $n_{mn}$ | $op_{mn}$ | $s_{m1}$ |

The algorithm used to project the StD relation $StDR_{system} \subseteq \Psi_1 \times N \times \Lambda \times \Psi_2$ from the ITG relation $ITGR_{system} \subseteq \Psi_1 \times N \times \Xi \times \Lambda \times \Theta \times \Gamma \times \Psi_2$ is as follows.

**ALGORITHM 2** (Projecting $StDR_{system}$ from $ITGR_{system}$)

**For** i = 1, m **Loop**

SELECT $\Psi_1$, $N$, $\Lambda$, $\Psi_2$ INTO $StDR_i$ FROM $ITGR_i$ ;

**End Loop;**

ORTHOGONALLY COMPOSE ALL $StDR_i$ (i.e., $\|_{i=1,m} StDR_i$) TO GET $StDR_{system}$

**END ALGORITHM**

Once we have the StD relation $StDR_{system}$, it is easy to get a UML 2.0 state diagram of the software system.

*F. Projecting the UML Sequence Diagram from the SBC Interaction Transition Graph*

In UML 2.0, the sequence diagram describes how a group of objects collaborate in some behavior - typically a single behavior. The diagrams show a number of example objects and the messages that are passed between these objects within the use-case. The notion of a UML 2.0 sequence diagram is defined as follows.

**DEFINITION** (SEQUENCE DIAGRAM) A UML sequence diagram $SqD = (E, N, \Xi, \Lambda, \Theta, \Gamma, SqDR)$ *consists of*

- *a finite set E of execution orders,*
- *a finite set N of operation call or operation return tags,*
- *a finite set $\Xi$ of external environment's actors or objects,*
- *a finite set $\Lambda$ of operation names,*
- *a finite set $\Theta$ of parameter lists,*
- *a finite set $\Gamma$ of objects,*
- *a relation $SqDR \subseteq E \times N \times \Xi \times \Lambda \times \Theta \times \Gamma$, where $(e, n, \rho, op, p, b) \in SqDR$.*

The UML 2.0 sequence diagram of a software system $SqD_{system}$ is defined as $\|_{i=1,m} SqD_i$ or $SqD_1 \| SqD_2 \| \ldots \| SqD_m$. Each sequence diagram $SqD_i$ is represented by a relation $SqDR_i \subseteq E \times N \times \Xi \times \Lambda \times \Theta \times \Gamma$, where $(e, n, \rho, op, p, b) \in SqDR_i$. The sequence diagram of a software system $SqD_{system}$ is represented by the relation $SqDR_{system}$ which is defined as $\|_{i=1,m} SqDR_i$ or $SqDR_1 \| SqDR_2 \| \ldots \| SqDR_m$, as shown in Table 5.

TABLE 5. Relation $SqDR_{system}$

| $E$ | $N$ | $\Xi$ | $\Lambda$ | $\Theta$ | $\Gamma$ |
|---|---|---|---|---|---|
| $e_{11}$ | $n_{11}$ | $\rho_{11}$ | $op_{11}$ | $p_{11}$ | $b_{11}$ |
| $e_{12}$ | $n_{12}$ | $\rho_{12}$ | $op_{12}$ | $p_{12}$ | $b_{12}$ |
| • | • | • | • | • | • |
| $e_{1n}$ | $n_{1n}$ | $\rho_{1n}$ | $op_{1n}$ | $p_{1n}$ | $b_{1n}$ |

$\|$

| $E$ | $N$ | $\Xi$ | $\Lambda$ | $\Theta$ | $\Gamma$ |
|---|---|---|---|---|---|
| $e_{21}$ | $n_{21}$ | $\rho_{21}$ | $op_{21}$ | $p_{21}$ | $b_{21}$ |
| $e_{22}$ | $n_{22}$ | $\rho_{22}$ | $op_{22}$ | $p_{22}$ | $b_{22}$ |
| • | • | • | • | • | • |
| $e_{2n}$ | $n_{2n}$ | $\rho_{2n}$ | $op_{2n}$ | $p_{2n}$ | $b_{2n}$ |

$\| \bullet \bullet \bullet \bullet \bullet \|$

| $E$ | $N$ | $\Xi$ | $\Lambda$ | $\Theta$ | $\Gamma$ |
|---|---|---|---|---|---|
| $e_{m1}$ | $n_{m1}$ | $\rho_{m1}$ | $op_{m1}$ | $p_{m1}$ | $b_{m1}$ |
| $e_{m2}$ | $n_{m2}$ | $\rho_{m2}$ | $op_{m2}$ | $p_{m2}$ | $b_{m2}$ |
| • | • | • | • | • | • |
| $e_{mn}$ | $n_{mn}$ | $\rho_{mn}$ | $op_{mn}$ | $p_{mn}$ | $b_{mn}$ |

The algorithm used to project the SqD relation $SqDR_{system} \subseteq E \times N \times \Xi \times \Lambda \times \Theta \times \Gamma$ from the ITG relation $ITGR_{system} \subseteq \Psi_1 \times N \times \Xi \times \Lambda \times \Theta \times \Gamma \times \Psi_2$ is as follows.

**ALGORITHM 3** (Projecting $SqDR_{system}$ from $ITGR_{system}$)

**For** i = 1, m **Loop**

CREATE RELATION $SqDR_i$ ($E$ int IDENTITY(1,1), $N, \Xi, \Lambda, \Theta, \Gamma$);

INSERT INTO $SqDR_i$ ($N, \Xi, \Lambda, \Theta, \Gamma$) SELECT $N, \Xi, \Lambda, \Theta, \Gamma$ FROM $ITGR_i$ ;

**End Loop;**

ORTHOGONALLY COMPOSE ALL $SqDR_i$ (i.e., $\|_{i=1,m} SqDR_i$) TO GET $SqDR_{system}$

**END ALGORITHM**

Once we have the SqD relation $SqDR_{system}$, it is easy to get a UML 2.0 sequence diagram of the software system.

## IV. CASE: ONLINE SHOPPING SYSTEM

### A. Online Shopping System

The online shopping system is a highly distributed world wide web-based software system that provides services for purchasing items such as books or clothes. In the online shopping system, customers can request to order one or more items from the supplier. The customer provides personal details, such as address and credit card information. This information is stored in a customer account. If the credit card is valid, then a delivery order is created and sent to the supplier. The supplier checks the available inventory, confirms the order, and enters a planned shipping date. When the order is shipped, the customer is notified and the customer's credit card account is charged. The online shopping system also allows the customer to view the details of the delivery order.

### B. SBC Interaction Transition Graph of the Online Shopping System

In SBC process algebra, the semantics of the online shopping system (OSS) is represented by a SBC interaction transition graph $ITG_{OSS}$ (defined as "$ITG_1 \| ITG_2 \| ITG_3$") with the transition relation $ITGR_{OSS} \subseteq \Psi_1 \times N \times \Xi \times \Lambda \times \Theta \times \Gamma \times \Psi_2$ (defined as "$ITGR_1 \| ITGR_2 \| ITGR_3$") as shown in Table 6.

TABLE 6. Relation $TGR_{OSS}$

| $S_1$ | $N$ | $\Xi$ | $\Lambda$ | $\Theta$ | $\Gamma$ | $S_2$ |
|---|---|---|---|---|---|---|
| $s_{11}$ | CAL | Customer | Request_Order_from_Customer | in Request_Order_Info | :Customer_UI | $s_{12}$ |
| $s_{12}$ | CAL | :Customer_UI | Request_Order_from_UI | in Request_Order_Info | :Customer_Coordinator | $s_{13}$ |
| $s_{13}$ | CAL | :Customer_Coordinator | Authorize_Credit_Card_Charge | in Credit_Card_Id; in Amount; out Authorization_Response | :Credit_Card_Service | $s_{14}$ |
| $s_{14}$ | CAL | :Customer_Coordinator | Store_Order | in Order; out Order_Id | :Delivery_Order_Service | $s_{15}$ |
| $s_{15}$ | RET | :Customer_UI | Request_Order_from_UI | out Order_Info | :Customer_Coordinator | $s_{16}$ |
| $s_{16}$ | RET | Customer | Request_Order_from_Customer | out Order_Info | :Customer_UI | $s_{11}$ |

$\|$

| $S_1$ | $N$ | $\Xi$ | $\Lambda$ | $\Theta$ | $\Gamma$ | $S_2$ |
|---|---|---|---|---|---|---|
| $s_{21}$ | CAL | Supplier | Shipping | in Order_Id | :Supplier_UI | $s_{22}$ |
| $s_{22}$ | CAL | :Supplier_UI | Ready_for_Shippment | in Order_Id | :Supplier_Coordinator | $s_{23}$ |
| $s_{23}$ | CAL | :Supplier_Coordinator | Request_Invoice | in Order_Id; out Invoice | :Delivery_Order_Service | $s_{24}$ |
| $s_{24}$ | CAL | :Supplier_Coordinator | Commit_Credit_Card_Charge | in Credit_Card_Id; in Amount; out Commit_Response | :Credit_Card_Service | $s_{25}$ |
| $s_{25}$ | CAL | :Supplier_Coordinator | Confirm_Payment | in Credit_Order_Id; in Amount; out Order_Status | :Delivery_Order_Service | $s_{26}$ |
| $s_{26}$ | RET | :Supplier_UI | Ready_for_Shippment | out Order_Status | :Supplier_Coordinator | $s_{27}$ |
| $s_{27}$ | RET | Supplier | Shipping | out Order_Status | :Supplier_UI | $s_{21}$ |

$\|$

TABLE 6 (continued). Relation $TGR_{OSS}$

| $S_1$ | $N$ | $\Xi$ | $\Lambda$ | $\Theta$ | $\Gamma$ | $S_2$ |
|---|---|---|---|---|---|---|
| $s_{31}$ | CAL | Customer | Request_Order_Status_from_Customer | in Order_Id | :Customer_UI | $s_{32}$ |
| $s_{32}$ | CAL | :Customer_UI | Request_Order_Status_from_UI | in Order_Id | :Customer_Coordinator | $s_{33}$ |
| $s_{33}$ | CAL | :Customer_Coordinator | Read_Order | in Order_Id; out Order | :Delivery_Order_Service | $s_{34}$ |
| $s_{34}$ | RET | :Customer_UI | Request_Order_Status_from_UI | out Order_Status | :Customer_Coordinator | $s_{35}$ |
| $s_{35}$ | RET | Customer | Request_Order_Status_from_Customer | out Order_Status | :Customer_UI | $s_{31}$ |

## C. Projecting the UML Class Diagram of the Online Shopping System

We apply the algorithm of projecting the ClsD relation (i.e., $ClsDR_{OSS}$) from the ITG relation (i.e., $ITGR_{OSS}$) of the online shopping system. After the projection, we get the relation $ClsDR_{OSS} \subseteq C \times \Lambda \times \Theta$ as shown in Table 7.

TABLE 7. Relation $ClsDR_{OSS}$

| $C$ | $\Lambda$ | $\Theta$ |
|---|---|---|
| :Customer_UI | Request_Order_from_Customer | in Request_Order_Info; out Order_Info |
| :Customer_UI | Request_Order_Status_from_Customer | in Order_Id; out Order_Status |
| :Supplier_UI | Shipping | in Order_Id; out Order_Status |
| :Customer_Coordinator | Request_Order_from_UI | in Request_Order_Info; out Order_Info |
| :Customer_Coordinator | Request_Order_Status_from_UI | in Order_Id; out Order_Status |
| :Supplier_Coordinator | Ready_for_Shippment | in Order_Id |
| :Credit_Card_Service | Authorize_Credit_Card_Charge | in Credit_Card_Id; in Amount; out Authorization_Response |
| :Credit_Card_Service | Commit_Credit_Card_Charge | in Credit_Card_Id; in Amount; out Commit_Response |
| :Delivery_Order_Service | Store_Order | in Order; out Order_Id |
| :Delivery_Order_Service | Request_Invoice | in Order_Id; out Invoice |
| :Delivery_Order_Service | Confirm_Payment | in Credit_Order_Id; in Amount; out Order_Status |
| :Delivery_Order_Service | Read_Order | in Order_Id; out Order |

From the projected ClsD relation $ClsDR_{OSS}$, we draw the corresponding UML 2.0 class diagram of the online shopping system, as shown in Figure 4.

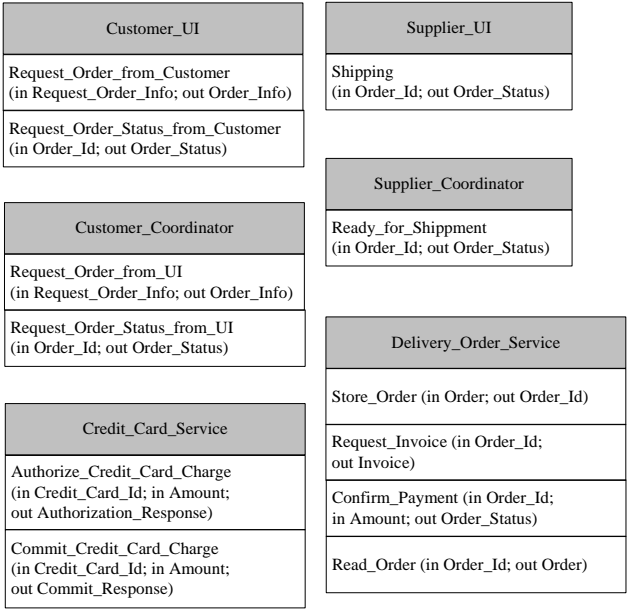

Figure 4. Projected ClsD View of the Online Shopping System

*D. Projecting the UML State Diagram of the Online Shopping System*

We apply the algorithm of projecting the StD relation (i.e., $StDR_{OSS}$) from the ITG relation (i.e., $ITGR_{OSS}$) of the online shopping system. After the projection, we get the relation $StDR_{OSS} \subseteq \Psi_1 \times N \times \Lambda \times \Psi_2$ as shown in Table 8.

TABLE 8. Relation $StDR_{OSS}$

| $\Psi_1$ | $N$ | $\Lambda$ | $\Psi_2$ |
|---|---|---|---|
| $s_{11}$ | CAL | Request_Order_from_Customer | $s_{12}$ |
| $s_{12}$ | CAL | Request_Order_from_UI | $s_{13}$ |
| $s_{13}$ | CAL | Authorize_Credit_Card_Charge | $s_{14}$ |
| $s_{14}$ | CAL | Store_Order | $s_{15}$ |
| $s_{15}$ | RET | Request_Order_from_UI | $s_{16}$ |
| $s_{16}$ | RET | Request_Order_from_Customer | $s_{11}$ |

| $\Psi_1$ | $N$ | $\Lambda$ | $\Psi_2$ |
|---|---|---|---|
| $s_{21}$ | CAL | Shipping | $s_{22}$ |
| $s_{22}$ | CAL | Ready_for_Shippment | $s_{23}$ |
| $s_{23}$ | CAL | Request_Invoice | $s_{24}$ |
| $s_{24}$ | CAL | Commit_Credit_Card_Charge | $s_{25}$ |
| $s_{25}$ | CAL | Confirm_Payment | $s_{26}$ |
| $s_{26}$ | RET | Ready_for_Shippment | $s_{27}$ |
| $s_{27}$ | RET | Shipping | $s_{21}$ |

| $\Psi_1$ | $N$ | $\Lambda$ | $\Psi_2$ |
|---|---|---|---|
| $s_{31}$ | CAL | Shipping | $s_{32}$ |
| $s_{32}$ | CAL | Ready_for_Shippment | $s_{33}$ |
| $s_{33}$ | CAL | Request_Invoice | $s_{34}$ |
| $s_{34}$ | RET | Ready_for_Shippment | $s_{35}$ |
| $s_{35}$ | RET | Shipping | $s_{31}$ |

From the projected StD relation $StDR_{OSS}$, we draw the corresponding UML 2.0 state diagram of the online shopping system, as shown in Figure 5.

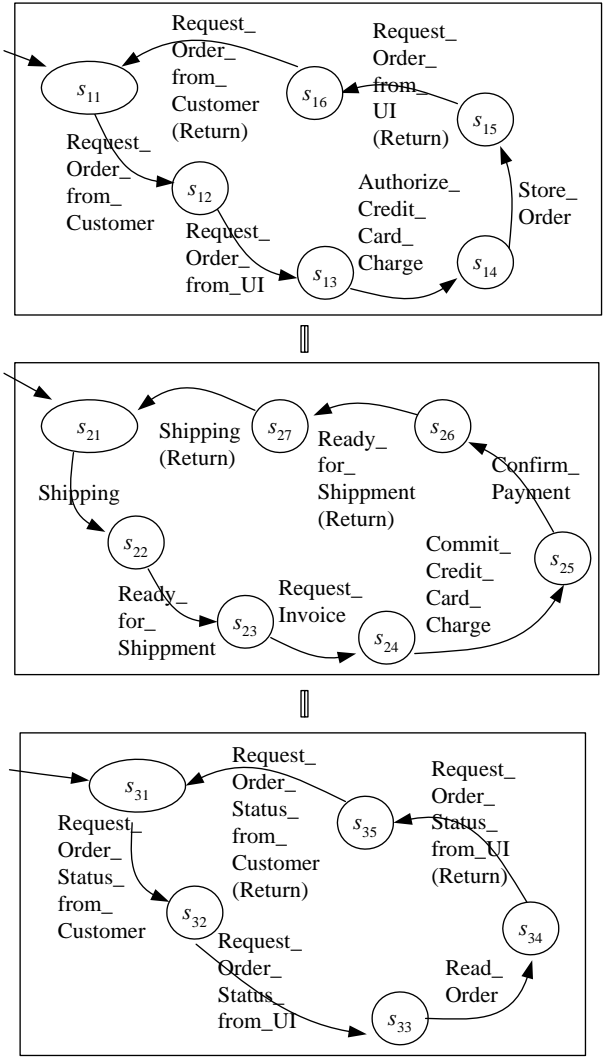

Figure 5. Projected StD View of the Online Shopping System

### E. Projecting the UML Sequence Diagram of the Online Shopping System

We apply the algorithm of projecting the SqD relation (i.e., $SqDR_{OSS}$) from the ITG relation (i.e., $ITGR_{OSS}$) of the online shopping system. After the projection, we get the relation $SqDR_{OSS} \subseteq E \times N \times \Xi \times \Lambda \times \Theta \times \Gamma$ as shown in Table 9.

TABLE 9. Relation $SqDR_{OSS}$

| $E$ | $N$ | $\Xi$ | $\Lambda$ | $\Theta$ | $\Gamma$ |
|---|---|---|---|---|---|
| 1 | CAL | Customer | Request_Order_from_Customer | in Request_Order_Info | :Customer_UI |
| 2 | CAL | :Customer_UI | Request_Order_from_UI | in Request_Order_Info | :Customer_Coordinator |
| 3 | CAL | :Customer_Coordinator | Authorize_Credit_Card_Charge | in Credit_Card_Id; in Amount; out Authorization_Response | :Credit_Card_Service |
| 4 | CAL | :Customer_Coordinator | Store_Order | in Order; out Order_Id | :Delivery_Order_Service |
| 5 | RET | :Customer_UI | Request_Order_from_UI | out Order_Info | :Customer_Coordinator |
| 6 | RET | Customer | Request_Order_from_Customer | out Order_Info | :Customer_UI |

⇓

| $E$ | $N$ | $\Xi$ | $\Lambda$ | $\Theta$ | $\Gamma$ |
|---|---|---|---|---|---|
| 1 | CAL | Supplier | Shipping | in Order_Id | :Supplier_UI |
| 2 | CAL | :Supplier_UI | Ready_for_Shippment | in Order_Id | :Supplier_Coordinator |
| 3 | CAL | :Supplier_Coordinator | Request_Invoice | in Order_Id; out Invoice | :Delivery_Order_Service |
| 4 | CAL | :Supplier_Coordinator | Commit_Credit_Card_Charge | in Credit_Card_Id; in Amount; out Commit_Response | :Credit_Card_Service |
| 5 | CAL | :Supplier_Coordinator | Confirm_Payment | in Credit_Order_Id; in Amount; out Order_Status | :Delivery_Order_Service |
| 6 | RET | :Supplier_UI | Ready_for_Shippment | out Order_Status | :Supplier_Coordinator |
| 7 | RET | Supplier | Shipping | out Order_Status | :Supplier_UI |

⇓

TABLE 9 (continued).   Relation $SqDR_{OSS}$

| $E$ | $N$ | $\Xi$ | $\Lambda$ | $\Theta$ | $\Gamma$ |
|---|---|---|---|---|---|
| 1 | CAL | Customer | Request_Order_Status_from_Customer | in Order_Id | :Customer_UI |
| 2 | CAL | :Customer_UI | Request_Order_Status_from_UI | in Order_Id | :Customer_Coordinator |
| 3 | CAL | :Customer_Coordinator | Read_Order | in Order_Id; out Order | :Delivery_Order_Service |
| 4 | RET | :Customer_UI | Request_Order_Status_from_UI | out Order_Status | :Customer_Coordinator |
| 5 | RET | Customer | Request_Order_Status_from_Customer | out Order_Status | :Customer_UI |

From the projected SqD relation $SqDR_{OSS}$, we draw the corresponding UML 2.0 sequence diagram of the online shopping system, as shown in Figure 6.

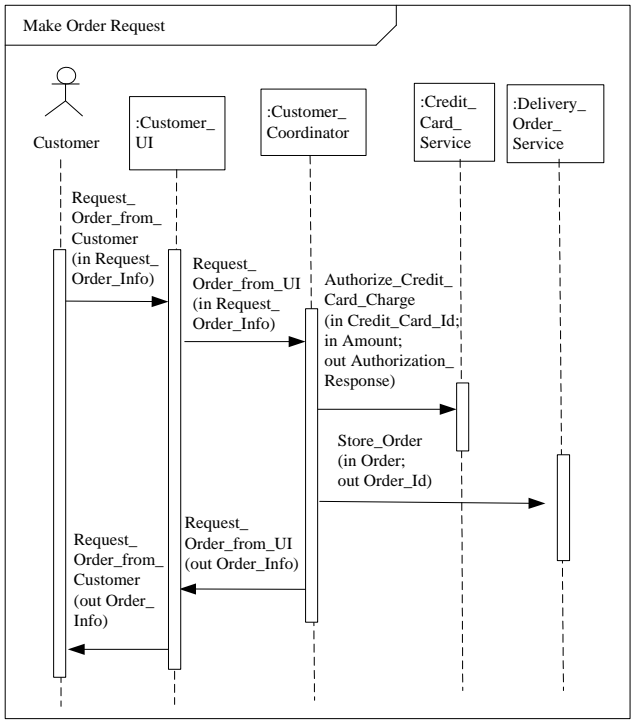

Figure 6.  Projected SqD View of the Online Shopping System  (I)

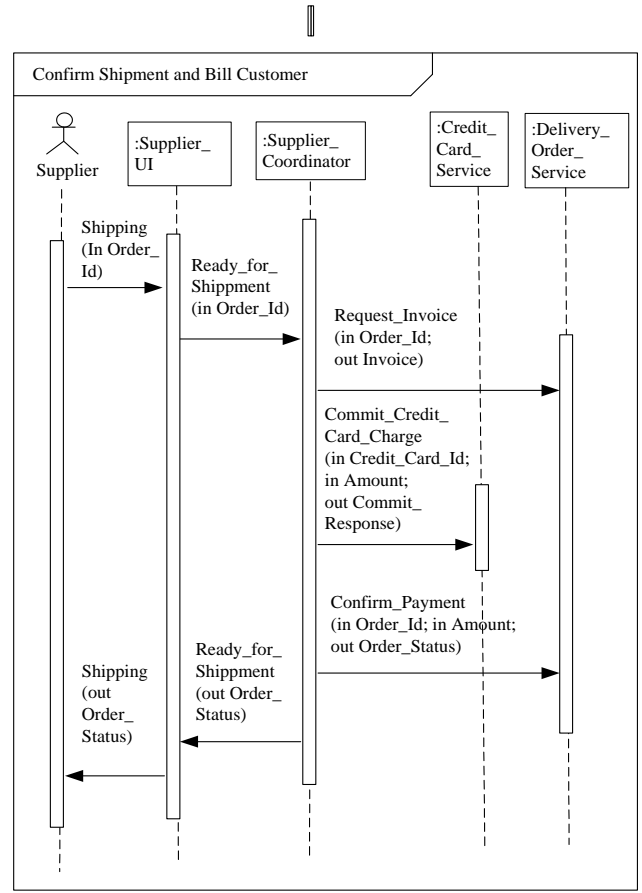

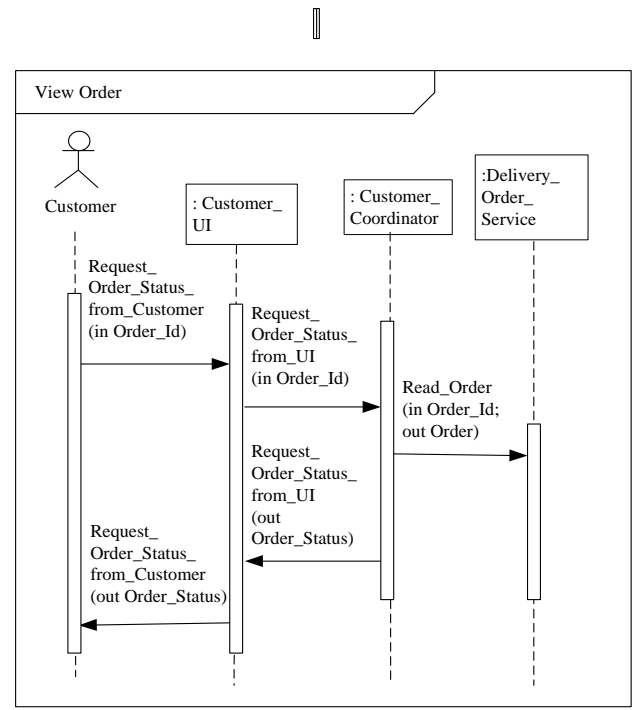

Figure 6.  Projected SqD View of the Online Shopping System  (II)

## V. CONCLUSIONS

In this paper, Operation-Based Multi-Queue Structure-Behavior Coalescence Process Algebra (O-M-SBC-PA) is proposed as a metamodel for UML 2.0 in model-driven engineering. One important use of the UML 2.0 metamodel is to provide an integrated semantic framework that every diagram in the user model can be projected as a view of the metamodel. Nowadays, most current UML 2.0 metamodels fail to project each diagram in the user model as a view of the metamodel.

In order to overcome the shortcomings of the current UML 2.0 metamodel approach, we need an integrated semantic framework that is able to unify structural constructs and behavioral constructs. Adopting O-M-SBC-PA as a metamodel for UML 2.0, we use the SBC transition graph as a single diagram to complete the overall semantic specification of the software system. Through the SBC transition graph and its corresponding transition relation, each diagram in the UML 2.0 user model can be projected as a view of the SBC transition graph. Therefore, we conclude that the SBC transition graph used by the SBC method as a metamodel for UML 2.0 is indeed a basis for unification of different views of the software system in Model-Driven Engineering.

## ACKNOWLEDGMENTS

The author wishes to express his thanks to the anonymous references for their valuable comments, which help clarify subtle points and triggered new ideas.

**William S. Chao** was born in 1954 in Taiwan and received his Ph.D. degree in information science from the University of Alabama at Birmingham, USA, in 1988. William worked as a computer scientist at GE Research and Development Center, from 1988 till 1991 and has been teaching at National Sun Yat-Sen University, Taiwan since 1992. His research covers: systems architecture, hardware architecture, software architecture, and enterprise architecture. Dr. Chao is a member of the Association of Enterprise Architects Taiwan Chapter and also a member of the Chinese Association of Enterprise Architects